\documentstyle[prl,aps,epsf]{revtex}

\begin{document}
\draft

\twocolumn[\hsize\textwidth\columnwidth\hsize\csname  @twocolumnfalse\endcsname 
\title{Critical Dynamics of Burst Instabilities in the Portevin-Le Ch\^{a}telier Effect}
\author{Gianfranco D'Anna$^{1}$ and Franco Nori$^{2}$}
\address{
$^1$ Institut de G\'{e}nie Atomique, Ecole Polytechnique F\'{e}d\'{e}rale de
Lausanne, CH-1015 Lausanne, Switzerland}
\address{
$^2$ Department of Physics, The University of Michigan, Ann Arbor, Michigan 48109-1120}
\date{\today}
\maketitle

\begin{abstract}
We investigate the Portevin-Le Ch\^{a}telier effect (PLC),
by compressing Al-Mg alloys in a very large deformation range, 
and interpret the results from the viewpoint of phase 
transitions and critical phenomena.
The system undergoes two dynamical phase transitions between 
intermittent (or ``jerky") and ``laminar" plastic dynamic phases.
Near these two dynamic critical points, the order parameter $1/\tau$ 
of the PLC effect exhibits large fluctuations, and 
``critical slowing down" (i.e., the number $\tau$ of bursts, or 
plastic instabilities, per unit time slows down considerably).
%
%
\end{abstract}
\vspace{-5pt}
\pacs{PACS numbers: 62.20.Fe, 81.40.Lm, 83.50.-v}
\vspace{-25pt}
\vskip2pc]

The motion of lattice dislocations is the main source of 
macroscopic plastic deformation in crystalline solids. 
In principle, the motion of a single dislocation in 
an otherwise perfect crystal is a simple phenomenon.  
However, in a real crystal, many cooperative 
and core dislocation effects 
may give rise to unusual and complex plastic 
properties\cite{FridelGeneral}.
An intriguing example of the latter is provided by the 
Portevin-Le Ch\^{a}telier (PLC) effect \cite{Portevin}, 
as observed in physical metallurgy (see, e.g., 
\cite{PLC review,MertemsPRL97}). 
In stress-strain experiments in which the deformation rate 
is held constant, the stress usually increases monotonically 
with the strain, and the solid deforms homogeneously in a 
continuously flowing or ``laminar'' plastic regime.
When the PLC effect arises, notably in some alloys and
at high deformation, intermittent yielding points, 
or {\it plastic instabilities\/} are observed, and the 
deformation is spatially inhomogeneous and apparently
confined to mesoscopic channels in the sample. 
This intermittently flowing or ``jerky'' plastic regime 
has been analyzed in terms of microscopic ageing 
models, which ascribe it to {\it dislocation-avalanches\/} 
triggered by the interaction of dislocations and clouds 
of mobile impurities\cite{Cottrell}, or to closely related 
mechanisms (see, e.g., \cite{PLC review,VariousTheo}). 
In these models, mobile point defects are attracted by the
stress field of an immobile dislocation, condensing in a 
cloud around the dislocation. Thus, the pinning force 
necessary to move the dislocation increases in time and 
the system ages.  By increasing the applied force, 
and as soon as the stress becomes large enough, 
the pinning breaks down and the dislocation 
flashes over a large distance to stop again. 
This distance is determined by the arrangement of the other 
dislocations sliding along the same channel and by strong 
pinning defects. This {\it cooperative\/} process repeats 
itself again and again, producing a succession of plastic 
instabilities.

Beyond its importance in metallurgy, the PLC 
effect is a paradigm for a general class of nonlinear 
complex systems with intermittent bursts.
Indeed, the succession of plastic instabilities shares 
both physical and statistical properties with many 
other systems exhibiting loading-unloading cycles. 
This is because the macroscopic statistical properties are 
a consequence of similar underlying microscopic physical 
mechanisms (e.g., log-interacting dislocation lines or 
magnetic flux lines, slowly driven in a disordered landscape 
with pinning traps, and obeying glassy dissipative dynamics).
Weertman\cite{Weertman}, and more recently 
Lebyodkin {\it et al}.\cite{LebyodkinPRL95}, have 
analyzed the analogy of the PLC effect with earthquakes.
But similarities also exist to stick-slip phenomena 
in sheared granular media\cite{stickgranular}, avalanches of 
magnetic vortices in superconductors\cite{FieldAvalancPRL95}, 
``starquakes'' with $\gamma$--ray burst activity in magnetars\cite{ChengNature96}, 
and even bursts of economic activity and stock market crashes\cite{EconoBursts}.
The dynamic behavior of some of these systems can be viewed in 
terms of dynamic phases separated by dynamical critical points 
(see, e.g., \cite{dynamic_phases} and refs.~therein). 
This is achieved by adapting concepts of equilibrium phase 
transitions \cite{critical_phenomena} to non-equilibrium cases.
It is tempting to extend the analogy to the PLC effect 
by further studying the dynamics of the jerky regime.
Here we present results of compression experiments on $34$
polycrystalline Al-Mg samples with nominal 4\%Mg. 
In contrast to the more common traction technique, 
compression experiments allow to access very large deformations 
without sample failure, and to collect very long time-series.
We interpret our results from the viewpoint of phase 
transitions and critical phenomena.

Experiments were conducted on $34$ samples of the alloy Al-4at.\%Mg; 
a classical system exhibiting the PLC effect\cite{PLC review,LebyodkinPRL95}. 
Data were collected at room temperature, about 18${{}^\circ}C$. 
Samples were slightly oblique cylinders of about 6~mm length
and 4~mm diameter, inclined 2 degrees from the vertical;
Grain size in the virgin samples was about 100~$\mu$m. 
In Fig.~1 we show a typical segment of a stress versus time curve, 
for an imposed compression speed $v=0.18~\mu $m/s, and 
when the sample is in the range of a 17\% deformation. 
(The compression velocity is $v=dl/dt$, where $l$ is the actual
length of the sample. Deformation is calculated assuming constant volume by $%
\epsilon =-\ln \left( 1-\Delta l/l_{o}\right) $, where $\Delta l$ is the
elongation and $l_{o}$ the initial length of the sample \cite{Poirier}). 
One sees various sequences of elastic reloading of the sample until a yield
point is reached, followed by the sharp stress drop typical of the PLC effect. 
In Fig.~1 we show the negative derivative of the stress versus time, 
$-d\sigma /dt$. The latter evidences the stress drops as sharp ``bursts'' 
of energy and permits us to clearly define the waiting time $\tau _{n}$ 
between successive events, and the size $s_{n}$ of the nth burst.
In this way, we also eliminate difficulties related to the small drift of the average 
stress with the strain hardening, which results in an almost constant offset in $-d\sigma /dt$.

FIG. 1. Short time-sequence, taken in the range of $17\%$ deformation, 
of the stress $\sigma$ (and also $-d\sigma /dt$) versus time
for an Al-4at.\%Mg polycrystalline sample.  For all figures, the 
imposed compression speed $v$ is $0.18~\mu $m/s, unless otherwise 
indicated.


FIG. 2. Various $-d\sigma /dt$ versus time for short 
time-sequences taken from a single measurement over a long time. 
(a) The first plastic instability appears at about 16\% deformation. 
(b)-(c) The average time $\tau$ between successive plastic instabilities 
diminishes when increasing the deformation. 
(d)-(e) For large deformations, $\tau$ increases and becomes more periodic.

The burst history of a sample is summarized in Fig.~2, which shows five
segments of $-d\sigma /dt$ versus time, taken along a single stress versus
strain curve measured at constant compression speed $v=0.18~\mu $m/s. 
The first burst appears at about 16\% deformation (about 6000 sec after the
beginning of the experiment). Above this lower critical deformation, in Fig. 2(a), 
we observe intense bursts which come in groups of one, two, two, three, and so on. 
We follow this jerky dynamic regime up to very large deformations. When
increasing the deformation, $\tau _{n}$ initially tends to diminish, as the number
of events per unit time increases (Fig. 2(b,c)); but beyond about 35\% deformation, 
$\tau_{n}$ tends to increase (Fig. 2(d,e)). At low deformations, below 35\%, 
the sequence of bursts seems random, possibly chaotic \cite{PLC review,Anan}. 
For large compressions, the sequence of bursts develops 
a periodic component (Fig. 2(d,e)). We do not observe here the
``Gutenberg-Richter'' power law in the distribution of burst sizes, as in
ref.~\cite{LebyodkinPRL95}, but a bell-shaped distribution instead.

The variations of the waiting time $\tau _{n}$ between successive events
with the deformation, $\varepsilon _{n},$ is shown in Fig.~3. 
These data (as well as other measurements in different samples at 
similar compression speeds) show that $\tau _{n}$ {\it diverges}, at both 
low and high deformations, according to power-laws of the form 
$\tau _{n}=A_{1}\left( \epsilon _{n}-\epsilon_{c1}\right) ^{-\beta _{1}}$, and 
$\tau _{n}=A_{2}\left( \epsilon_{c2}-\epsilon _{n}\right) ^{-\beta _{2}}$. 
For the divergence at small deformations, we find 
$\epsilon _{c1}$=0.13 and $\beta _{1}$=0.58. 
The second divergence occurs at very large deformations, with 
$\epsilon _{c2}$=1.1 and $\beta _{2}$=1.8.
In Fig.~3 we show that the system undergoes transitions 
between jerky and laminar dynamic phases, and exhibits 
large {\it fluctuations\/} and {\it critical slowing down\/} 
in the number of bursts per unit time in the jerky phase 
close to two non-equilibrium {\it dynamic critical points}. 
The divergences are of the form 
 $\tau \propto \left| \epsilon - \epsilon_{ci} \right|^{-\beta_i}$;
%
%
where $t_{i} = \left| \epsilon - \epsilon_{ci} \right| $, ($i=1,2$), 
behave as the analog of the ``reduced temperature" for each one of 
the two phase boundaries separating the different dynamic phases.
Surprisingly, the system recovers the laminar plastic flow in 
the very large deformation regime.

The effect of the compression speed $v$ on the jerky regime is analyzed in
Fig.~4, which shows four segments of $-d\sigma /dt$ versus time, taken at
four different compression speeds (in four different samples) at about 25\%
deformation. By increasing $v$, the time $\tau _{n}$ 
between events decreases roughly following a power law. 
This is shown in Fig.~5, where the local average 
$\left\langle \tau _{n}\right\rangle \equiv \tau$ is plotted versus $v$.
This average for $\tau_n$ is local, in the sense that it is taken over about 30
events around a given deformation. 
In Fig.~5 we show the local average $\left\langle \tau _{n}\right\rangle $ 
for 25\% and 60\% deformation, respectively.  Only as a guide to the eye, 
a power-law ($ \tau =Bv^{-\alpha }$) fit to the data is also shown. 
We find $\alpha \approx 0.8$ at both low and high deformation. 
Notice that the parameters describing the critical 
slowing down close to the two dynamic critical points 
do not seem to be affected by $v$. 
For example, for twice the compression velocity $v$ used in Fig.~3, 
we still obtain two dynamical critical points.
One for small deformations, with 
$\epsilon _{c1}$=0.17 and $\beta _{1}$=0.51, and a second 
divergence for very large deformations, with 
$\epsilon _{c2}$=1.07 and $\beta_{2}$=1.6.

Fast driving rates and large deformations are almost equivalent in their
effect on the PLC dynamics. Fast driving rates push the system into a more
periodic-response regime, as do large deformations, while small driving
rates favor an aperiodic (possibly random or chaotic) response. 
A qualitative microscopic explanation can be given.
For very small driving rates, the system has time to undergo complex
loading-unloading cycles in which a large number of plastic channels are
activated and deactivated. At fast driving rates, few percolation-like 
channels quickly dominate the dynamics and characteristic times appear. 
Similarly, as the system is squeezed at large deformations, only few 
channels remain active and the system acquires a periodic response.  
Similar behavior has been observed in the plastic instabilities and 
burst-like cooperative motion of ``magnetic flux lines" (instead of our 
``dislocation lines" here) over pinning centers\cite{FieldAvalancPRL95}.

The revolution in the general understanding of equilibrium critical 
phenomena was preceded by the gradual realization that apparently 
dissimilar and unrelated phenomena (e.g., chemical, magnetic, and 
superfluid transitions) shared some commonalities near critical points.
More recently, the focus has shifted to systems far from equilibrium 
or in metastable or steady states, and to the search for common 
behaviors and trends near their phase transitions.
In our samples, we apply a generalized force, the externally imposed stress, 
and the samples respond by generating intermittent bursts.
This reponse is quantified by the rate of burst generation, 
$1/\tau$, which can be seen as the order parameter for the PLC effect; 
the analog, e.g., of the magnetization $M$ in a magnet, 
or the density of paired electrons in superconductors.
In the PLC effect, the order parameter is a temporal average or ``current",
the number of bursts per unit time; instead of a spatial average typical 
for equilibrium systems.
Thus, the order parameter for the PLC dynamic phase behaves as 
$1/\tau \propto \left| \epsilon - \epsilon_{ci} \right| ^{\beta}$, 
with critical exponent $\beta$, as obtained in Fig.~3.
The relative deformation, $\left| \epsilon - \epsilon_{ci} \right|$,  
acts as the reduced temperature.
In other words, in our samples:
$1/\tau \neq 0$ in the jerky PLC phase, and $1/\tau=0$ in the laminar phase.
For magnets: $M \neq 0$ in the ordered ferromagnetic phase, and $M=0$ in the disordered phase.
%
%
Microscopic ordering of a spin is induced by the
cooperative alignment of the neighboring spins.
{\it Microscopic\/} motion of a dislocation is induced by the
cooperative motion of the neighboring dislocations,
in the PLC phase.
As clearly seen in Fig.~3, 
{\it fluctuations\/} in the order parameter $1/\tau$ 
increase near the two dynamic critical points, 
in analogy with equilibrium phase transitions.

That a {\it far\/} from equilibrium transition can be described by an 
order parameter, is already known for lasers\cite{critical_phenomena}. 
We add the PLC effect.
Like in the PLC effect, 
a laser is a driven system kept in a stationary state out of equilibrium.  
When the optical pump power crosses a threshold, 
a coherent electromagnetic (EM) wave is generated.  
%
%
In a laser, the internal EM field induces order in the form of 
temporally {\it correlated\/} emission events, while random decay 
events destroy correlations.
%
%
In our samples, temporally and spatially {\it correlated\/} 
cooperative jump events occur in the PLC phase 
(e.g., some nearby dislocations jump, while the rest do not move); 
while many small uncorrelated dislocations move in the laminar phase.
Near the critical point, the laser's relaxation time increases.  
In the PLC effect, also the relaxation time increases---in the 
sense that the average time $\tau$ between bursts increases.


In summary, we have studied the PLC effect in alloys, focusing 
on its critical behavior near its two dynamical phase boundaries 
with enhanced fluctuations and diverging relaxation times 
(i.e., ``critical slowing down" of the jerky activity).
In the large compression regime, the time between successive plastic
instabilities diverges, until finally the jerky regime stops. 
The discovery of critical behavior in the PLC effect is 
relevant for the general study of non-equilibrium driven  
systems with loading-unloading cycles, which produce plastic 
nonlinear instabilities and complex stick-slip dynamics.
We do not know to which extent this dynamic behavior is universal, 
and if it provides a key for predicting the future evolution of jerky 
nonlinear systems. However, our findings show an up to now unexplored
approach for the PLC, possibly valid for other systems which share 
similar dynamics, including the seismic activity of crustal faults 
and the starquakes in the solid crust of superdense stars.

We thank J. Martin and J. Bonneville for providing the 
testing machines, E. Giacometti for his help in using them, 
and M.~Bretz and C.~Kurdak for comments on the text. 
This work is supported by the Swiss National Science Foundation.  
FN acknowledges partial support from the UM Center for the Study 
of Complex Systems, and DOE contract No.~W-31-109-ENG-38.


FIG. 1. Short time-sequence, taken in the range of $17\%$ deformation, 
of the stress $\sigma$ (and also $-d\sigma /dt$) versus time
for an Al-4at.\%Mg polycrystalline sample.  For all figures, the 
imposed compression speed $v$ is $0.18~\mu $m/s, unless otherwise 
indicated.

FIG. 2. Various $-d\sigma /dt$ versus time for short 
time-sequences taken from a single measurement over a long time. 
(a) The first plastic instability appears at about 16\% deformation. 
(b)-(c) The average time $\tau$ between successive plastic instabilities 
diminishes when increasing the deformation. 
(d)-(e) For large deformations, $\tau$ increases and becomes more periodic.

FIG. 3. The time $\tau _{n}$ between successive instabilities, versus the
deformation, $\varepsilon _{n},$ and power-law fits shown by dashed lines.
Large {\it fluctuations\/} and {\it critical slowing down\/} 
are evident near the two dynamic critical points.

FIG. 4. Various $-d\sigma /dt$ versus time for short segments at four 
different compression speeds $v$ in different samples, at about the 
same deformation of $25\%$.

FIG. 5. The average time 
between bursts versus compression speed $v$, 
and power-law fits.
These fits have no bearing into the main results 
(e.g., large fluctuations and diverging relaxation times, 
critical slowing down, in the order parameter $1/\tau$ 
near the two dynamical critical points) because 
the compressions shown, $25\%$ and $60\%$,
are far from the critical points.

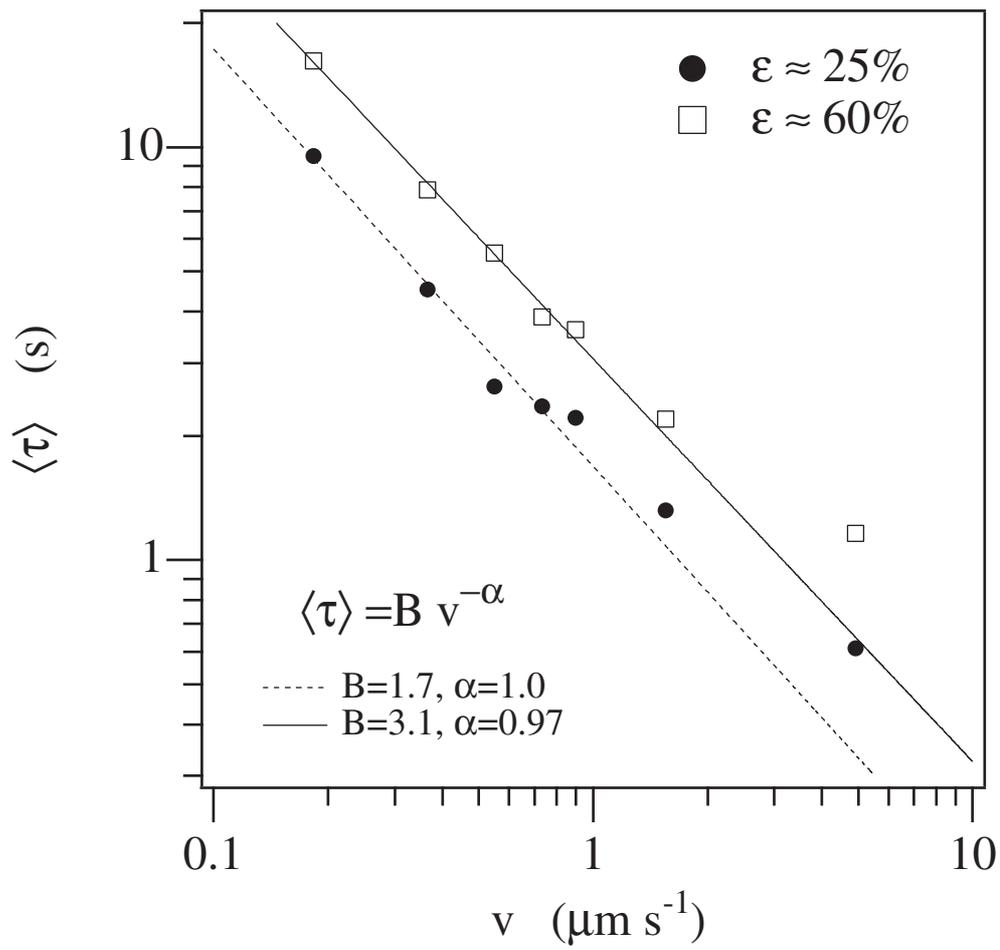
\end{document}